\documentstyle[12pt, epsf]{article}
\begin{document}
\begin{titlepage}
\title{
Dynamical mass generation of a $two$-component fermion in 
Maxwell-Chern-Simons $QED_3$: \\
The lowest ladder approximation
}
\author{
Toyoki Matsuyama$^1$ \\
Department of Physics, Nara University of Education \\
Takabatake-cho, Nara 630-8528, JAPAN \\
and \\
Hideko Nagahiro and Satoko Uchida \\
Department of Physics, Nara Women's University  \\
Nara 630-8506, JAPAN
}
\vspace{1cm}

\baselineskip=21.5pt

\maketitle

\begin{abstract}
Dynamical mass generation of a two-component fermion in $QED_3$ with 
a Chern-Simons term is investigated by solving the Schwinger-Dyson equation 
formulated in the lowest ladder approximation.  
Dependence of the dynamical fermion mass on a gauge-fixing parameter, 
a gauge coupling constant, and a topological mass is examined 
by approximated analytical and also numerical methods.  
The inclusion of the Chern-Simons term makes impossible to choose a peculiar 
gauge in which a wave function renormalization is absent.  
The numerical evaluation shows that the wave function renormalization is 
fairly close to 1 in the Landau gauge.  
It means that this gauge is still a specific gauge where the Ward-Takahashi 
identity is satisfied approximately.  
We also find that the dynamical mass is almost constant if the topological 
mass is larger than the coupling constant, while it decreases when the 
topological mass is comparable to or smaller than the coupling constant 
and tends to the value in $QED_3$ without the Chern-Simons term.  

\vskip 2cm

\end{abstract}

\footnotemark{e-mail address: matsuyat@nara-edu.ac.jp}

\end{titlepage}
\baselineskip=21.5pt
\parskip=3pt

\section*{I. Introduction}

In (2+1)-dimensions, generally in odd-dimensional space-time, a peculiar type 
of quantum field theory exists.  
That is a theory with a {\it Chern-Simons term} in its action.~\cite{SS,DJT}  
The Chern-Simons term may be added to the action by hand or induced in an 
effective action by a vacuum polarization effect of a fermion as the 
anomaly~\cite{RNSI}.  
Our understanding on the theory has been progressed especially through 
studies of the topological structure due to the Chern-Simons 
term.~\cite{TJZW}  

On the other hand, two important phenomena were discovered in condensed matter 
systems.  
One is the quantum Hall effect~\cite{QH} and the other is the high-$T_c$ 
superconductivity~\cite{HTC}.  
Both phenomena may be thought of as realization of macroscopic quantum 
effects in {\it planar} systems in which electrons have a strong 
correlation.  
It is challenging to ask whether and how the Chern-Simons term is related to 
these phenomena discovered in the two-dimensional electron systems.  
These situations have activated the study of the (2+1)-dimensional quantum 
field theories.  

In (2+1)-dimensional gauge theories, the behaviour in the infrared region 
seems to be unstable at least in the perturbative treatment.  
The Chern-Simons term gives the gauge field a topological mass without 
breaking gauge symmetry so that the term may rescue the theory 
from the infrared catastrophe.  
This is original motivation of including the Chern-Simons term in the 
action.~\cite{JTAPRS}  
Reminding those studies, we think that one of the most important problems 
which is still not clear is what is a role of the Chern-Simons term in a 
$nonperturbative$ region.  
It is our purpose to clarify how the Chern-Simons term affects  
nonperturbative dynamics.  

The dynamical mass generation of the fermion is an important phenomenon 
induced by the nonperturbative effects.  
In even-dimensional space-time, chiral symmetry forbids the mass of the 
fermion.  
The nonperturbative effects can break this symmetry giving the mass to the 
fermion.  
As is well known, we cannot define the chiral symmetry for the $two$-component 
fermion which belongs to the irreducible spinor representation in 
(2+1)-dimensions.  
Instead parity symmetry ($P_2$) forbids the two-component fermion having 
the mass.  
On the other hand, for the $four$-component fermion which is composed of $two$ 
two-component fermions, we can define a kind of "chiral" symmetry. 
The "chiral" transformation is defined by a combination of the parity and 
$Z_2$ flavour transformations. ($P_4=P \otimes Z_2$)  

Many studies on the dynamical mass generation in $QED_3$ have been done 
already by using the Schwinger-Dyson technique~\cite{PABCWABKW} or the 
lattice Monte-Carlo method~\cite{DKK}.  
In almost of previous works, these authors have studied the dynamical mass 
generation of the four-component fermions in the theory without the 
Chern-Simons term, which is a spontaneous symmetry breaking of $P_4$ but 
not $P_2$.  
These studies have been extended to the case with the Chern-Simons term 
but the fermion considered there has been the four-component 
one.~\cite{KKKM}  

The theory with $N$ four-component fermions is equivalent to the theory with 
$2N$ two-component fermions.  
In this sense, the theory with the four-component fermions can describe only 
the case with even number of flavours (counting the two-component fermions).  
It is interesting to ask what happens in the theory with odd number of 
flavours.  
This also is another motivation of the present work.  
The simplest version of the theory with odd number of flavours is the 
case with a single flavour.  
The first work concerned with the dynamical mass generation of the single 
$two$-component fermion in $QED_3$ without the Chern-Simons term was done 
in Ref. ~\cite{HMH}.  
The symmetry which was studied in the work is $P_2$ but not $P_4$.  
In this paper, we extend the work to the case with the Chern-Simons term.  

Thus we consider the dynamical mass generation of the $single$ 
$two$-component fermion in $QED_3$ $with$ the Chern-Simons term.  
The Schwinger-Dyson equations are formulated in the lowest ladder 
approximation, which are two coupled integral equations.  
Due to the Chern-Simons term, a drastic change happens in the structure of the 
Schwinger-Dyson equations.  
While the wave function renormalization is absent in the Landau gauge in 
$QED_3$ without the Chern-Simons term as far as we consider in the lowest 
ladder approximation, the inclusion of 
the Chern-Simons term makes impossible to choose a peculiar gauge in 
which the wave function is not renormalized.  
Because there are no {\it ad hoc} reasons that the Landau gauge is specific in 
$QED_3$ with the Chern-Simons term, we should study the coupled integral 
equations for various values of the gauge parameter.  

Our strategy in this work are as follows:  
First we solve the coupled integral equations in an approximated analytical 
method which is very crude but may be available for reference in a more 
complete numerical calculation.  
After that, the coupled integral equations are solved numerically.  
In these analyses, we study dependence of the dynamical mass on the 
gauge-fixing parameter, the coupling constant, and the topological mass.  

This paper is organized as follows:  
In Sec. II, we explain the model which we use.  
The Schwinger-Dyson equations in the lowest ladder approximation are derived 
in Sec. III.  
In Sec. IV, the equations are analyzed by the approximated analytical method. 
The numerical analysis of the equations are presented in Sec. V.  
We conclude our results with discussions in Sec. VI.  
In Appendices A and B, some useful formulae are summarized.  

\section*{II. Maxwell-Chern-Simons $QED_3$}

The model which we consider is an extended version of the usual 
three-dimensional quantum electrodynamics ($QED_3$).
It has both of the Maxwell and Chern-Simons terms as the action of the U(1) 
gauge field .  
We call this theory the Maxwell-Chern-Simons $QED_3$.~\cite{JTAPRS}  
The gauge field is coupled to the two-component Dirac fermion.  
The Lagrangian density of the theory is given by 
\begin{eqnarray}
{\cal L}= - \frac{1}{4} F_{\mu\nu} F^{\mu\nu}
          + \frac{\mu}{2} \varepsilon^{\mu \nu \rho} A_\mu \partial_\nu A_\rho
          - \frac{1}{2\alpha}(\partial_\mu A^\mu)^2
          + \bar{\psi}(i \not \! \partial - e \not \! \! A)\psi \ \ .
\label{lagrangian}
\end{eqnarray}
The second term in the right-hand side of Eq.(\ref{lagrangian}) is 
the so-called Chern-Simons term.  
It is well-known that the term gives the gauge field the mass $\mu$ without 
breaking the gauge symmetry.  
$\mu$ is called the topological mass because the Chern-Simons term 
has topological meaning as a secondary characteristic 
class~\cite{TJZW,CS}.  
$\alpha$ is the gauge-fixing parameter and $e$ is the gauge coupling 
constant.  
$\psi$ is the two-component fermion field which belongs to the irreducible 
spinor representation in (2+1)-dimensions.  
The Dirac matrices are defined by $\gamma^0=\sigma_3, \gamma^1=i\sigma_1, 
\gamma^2=i\sigma_2$ with $diag(g^{\mu\nu})=(1,-1,-1)$ where $\sigma_i$'s 
(i=1, 2, 3) are the Pauli matrices.  
The $\gamma^\mu$'s satisfy relations as 
$\{ \gamma^\mu, \gamma^\nu \}=2g^{\mu \nu}$, $\gamma^\mu \gamma^\nu = -i 
\epsilon^{\mu \nu \rho} \gamma_\rho + g^{\mu \nu}$ and $tr[\gamma^\mu 
\gamma^\nu ] = 2g^{\mu \nu}$.  
In this representation, there does not exist a matrix which anti-commutes 
with all of $\gamma^\mu$'s so that we cannot define the chiral 
transformation.  
This is a specific aspect of the odd-dimensional space-time.  
In even-dimensions, the chiral symmetry requires that a fermion is 
massless.  
In odd-dimensions, the chiral symmetry itself does not exist.  
Instead, the mass term of the fermion is forbidden by parity symmetry.  
The parity transformation is defined as 
\begin{eqnarray}
x=(t, x, y) &\rightarrow& x'=(t, -x, y) \ \ , \nonumber \\ 
\psi(x) &\rightarrow& \gamma^1 \psi(x') \ \ , \nonumber \\ 
A^0 (x) &\rightarrow& A^0 (x') \ \ , \label{parity} \\ 
A^1 (x) &\rightarrow& - A^1 (x') \ \ , \nonumber \\ 
A^2 (x) &\rightarrow& A^2 (x') \ \ .  \nonumber
\end{eqnarray}
Under the parity transformation, the mass term of the fermion and the 
Chern-Simons term change their signs.  
Thus the mass terms of both the fermion and the gauge field are forbidden by 
the parity symmetry.  
We study how the breaking of parity by the topological mass affects the 
mass generation of the fermion.

\section*{III. Schwinger-Dyson Equations}

As a non-perturbative method to evaluate the dynamical mass of the fermion, 
we use here the Schwinger-Dyson technique.  
The Schwinger-Dyson equation for the fermion self-energy $\Sigma(p)$ is 
written as
\begin{eqnarray}
\Sigma(p)=(-i e)^2 \int \frac{d^3k}{(2\pi)^3} \ \gamma^\mu \ 
  i S'_F (k) \ \Gamma^\nu(k,p-k) \ i D'_{\mu\nu}(p-k) \ \ .
\label{SDeqfull}
\end{eqnarray}
$\Gamma^\nu(k,p-k)$ is a full vertex function and $D'_{\mu\nu}(p-k)$ is 
a full propagator of the gauge field.  
$S'_F$ is the full propagator of the fermion field which is written as  
\begin{eqnarray}
i S'_F(p)=\frac{i}{A(p)\not \hspace{-0.8mm}p - B(p)}
        =\frac{i}{\not \hspace{-0.8mm}p-i\Sigma(p)} \ \ ,
\label{fermion}
\end{eqnarray}
where $A(p)$ and $B(p)$ are functions of $\sqrt{p_\mu p^\mu}$ in taking 
into account the relativistic nature, while $\Sigma(p)$ depends on $p_\mu$'s.  
$A(p)^{-1}$ is the wave function renormalization and $B(p)/A(p)$ is a 
mass induced by dynamical effects at the momentum scale $p$.  
The so-called dynamical mass $m_{phys}$ is defined by $m_{phys}=B(0)/A(0)$ 
as usual.  
It is useful to notice the relations as 
\begin{eqnarray}
tr[\Sigma(p)]=-2i B(p) \ , \ \ 
tr[\not \hspace{-0.8mm}p \Sigma(p)]=2i(A(p)-1)p^2 \ \ .
\label{relation}
\end{eqnarray}

To proceed our analysis of Eq.(\ref{SDeqfull}) further, we need to 
introduce any suitable approximation.  
In this paper, we limit ourselves to use the lowest ladder approximation 
where the full propagator of the gauge field and the full vertex are replaced 
by the free propagator and the tree vertex respectively as 
\begin{eqnarray}
i D'_{\mu\nu}(p-k) \approx i D_{\mu\nu}(p-k) \ , \ \ 
\Gamma^\nu(k,p-k) \approx \gamma^\nu \ \ .  
\label{ladder}
\end{eqnarray}
The analysis beyond the ladder approximation will appear elsewhere.   
The free propagator of the gauge field is derived from the Lagrangian density 
(\ref{lagrangian}) as 
\begin{eqnarray}
i D^{\mu\nu}(p)=-i\frac{1}{p^2-\mu^2} \left(g^{\mu\nu} 
       - \frac{p^\mu p^\nu}{p^2}\right)
       + \mu\frac{1}{p^2-\mu^2}\frac{1}{p^2}\varepsilon^{\mu\nu\rho}p_\rho
       - i\alpha\frac{p^\mu p^\nu}{p^4} \ \ .
\label{gauge}
\end{eqnarray}
Thus the Schwinger-Dyson equation in the lowest ladder approximation becomes 
\begin{eqnarray}
\Sigma(p)=(-i e)^2\int\frac{d^3k}{(2\pi)^3}
        \gamma^\mu \,i S'_F(k)\gamma^\nu \,i D_{\mu\nu}(p-k) \ \ .
\label{SDeqld}
\end{eqnarray}

We substitute Eqs.(\ref{fermion}) and (\ref{gauge}) into Eq.(\ref{SDeqld}) 
and use Eq.(\ref{relation}) in producing two coupled equations. 
After taking the traces, we obtain the coupled integral equations as 
\begin{eqnarray}
B(p) &=& - \frac{i e^2}{2} \int\frac{d^3k}{(2\pi)^3}  
           \frac{1}{ A(k)^2 k^2- B(k)^2} \left[-2\mu A(k) 
           \frac{p^2-k^2-(p-k)^2} {\{(p-k)^2-\mu^2\}(p-k)^2} \right. 
           \nonumber \\
       &+& \left. 4 B(k) \frac{1}{(p-k)^2-\mu^2} 
           +2\alpha B(k) \frac{1}{(p-k)^2}\right] \ \ , 
\label{Bd3k} \\
A(p) &=& 1- \frac{i e^2}{2 p^2} \int\frac{d^3k}{(2\pi)^3} 
            \frac{1}{ A(k)^2 k^2- B(k)^2}
            \left[  \frac{A(k)}{(p-k)^2-\mu^2}  
            \left\{\frac{(p^2-k^2)^2}{(p-k)^2}-(p-k)^2 \right\} \right.
            \nonumber \\
        &-& \frac{2\mu B(k)}{(p-k)^2-\mu^2} 
            \left\{\frac{p^2-k^2}{(p-k)^2}+1 \right\} 
          - \left. \alpha A(k) \left\{\frac{(p^2-k^2)^2}{(p-k)^4}
          - \frac{p^2+k^2}{(p-k)^2} \right\} \right] \ \ .
\label{Ad3k}
\end{eqnarray}

Now we change the metric to the Euclidean one by the Wick rotation as 
$(k^0,\vec{k}) \rightarrow (ik^0,\vec{k})$ and $(p^0,\vec{p}) \rightarrow 
(ip^0,\vec{p})$.  
Then $k^2$ and $p^2$ are replaced by $-k^2=-(k^0)^2-(k^1)^2$ and $-p^2=
-(p^0)^2-(p^1)^2$.  
After that,  we transform the integral variables $k^\mu$'s to the polar 
coordinates $(k, \theta,\phi)$.  
The angular integration on $\theta$ and $\phi$ can be done explicitly.  
(The useful integral formulae are listed in Appendix A.)
Finally we obtain the coupled integral equations which contain only the 
integration on the radial variable $k$,  
\begin{eqnarray}
B(p) &=& \frac{e^2}{8\pi^2p} \int_{0}^{\infty}dk 
            \frac{k}{A(k)^2 k^2+B(k)^2} \left[ \left\{\alpha B(k) - 
            \frac{1}{\mu}(p^2-k^2) A(k) \right\} 
            \ln\frac{(p+k)^2}{(p-k)^2} \right. \nonumber \\
        &+& \left. \left\{\frac{1}{\mu}(p^2-k^2) A(k) +\mu A(k) 
         +2 B(k) \right\} \ln\frac{(p+k)^2+\mu^2}{(p-k)^2+\mu^2} \right] 
         \ \ , 
\label{B} \\
A(p) &=& 1+\frac{e^2}{8\pi^2p^3}\int_{0}^{\infty}dk
            \frac{k}{A(k)^2 k^2+ B(k)^2} 
            \left[ -2pk(\alpha+1) A(k) 
          + \left\{\frac{1}{2\mu^2}(p^2-k^2)^2 A(k) 
            \right. \right. \nonumber \\
        &+& \left. \frac{1}{\mu}(p^2-k^2) B(k) 
          + \frac{1}{2}\alpha(p^2+k^2) A(k) \right\} 
            \ln\frac{(p+k)^2}{(p-k)^2}  
          + \left\{\frac{1}{2}\mu^2 A(k) \right. \nonumber \\
        &-& \left. \left. \frac{1}{2\mu^2}(p^2-k^2)^2 A(k)
           +\mu B(k) - \frac{1}{\mu}(p^2-k^2) B(k) \right\}
            \ln\frac{(p+k)^2+\mu^2}{(p-k)^2+\mu^2} \right]
            \ \ .
\label{A}
\end{eqnarray}
In the successive Secs. IV and V, we solve these equations by an approximated 
analytical method and also numerically by using an iteration method.  

Notice that the parameters $e^2$ and $\mu$ have the dimensions of mass.  
We may rewrite Eqs. (\ref{B}) and (\ref{A}) to dimensionless forms by 
defining $\hat{\mu} \equiv \mu/e^2$, $b(x) \equiv B(e^2 x)/e^2$ and 
$a(x) \equiv A(e^2 x)$ where $x$ is a dimensionless variable defined by 
$p=e^2 x$.  
The equations are just the ones obtained by setting $e^2=1$ and replacing 
$(A, B, \mu)$ to $(a, b, \hat{\mu})$ in Eqs. (\ref{B}) and (\ref{A}).  
The theory is controlled by only one dimensionless parameter $\hat{\mu}$.  
After solving the dimensionless equations for $a(x)$ and $b(x)$, we can 
convert them to $A(p)$ and $B(p)$.  

\section*{IV. Approximated Analytical Studies}

\subsection*{IV-A. $\mu \rightarrow 0$ limit}

We can check easily that Eqs. (\ref{B}) and (\ref{A}) reduce to the 
Schwinger-Dyson equations in $QED_3$ without the Chern-Simons term if we 
put the topological mass $\mu$ equal to zero.  
In fact, taking the limit as $\mu \rightarrow 0$ in Eqs. (\ref{B}) and 
(\ref{A}), we obtain 
\begin{eqnarray}
B(p)&=&(\alpha+2) \frac{e^2}{8 \pi^2 p} \int^\infty_0 dk \frac{kB(k)} 
     {A(k)^2 k^2 + B(k)^2} \ln\frac{(p+k)^2}{(p-k)^2} \ \ , 
\label{BwoCS} \\
A(p)&=&1 - \alpha \frac{e^2}{4 \pi^2 p^3} \int^\infty_0 dk \frac{kA(k)} 
     {A(k)^2 k^2 + B(k)^2} \left[ pk - \frac{p^2+k^2}{4} 
      \ln\frac{(p+k)^2}{(p-k)^2} \right] \ \ ,
\label{AwoCS}
\end{eqnarray}
which are the Schwinger-Dyson equations in the lowest ladder approximation 
derived in $QED_3$ without the Chern-Simons term.\footnote{In ~\cite{HMH}, 
there were some typing mistakes.  
We give the correct equations in Eqs. (\ref{BwoCS}) and (\ref{AwoCS}).}
We can see that there exists the specific gauge where the wave function 
renormalization is absent.  
Thus in the Landau gauge ($\alpha=0$), Eq. (\ref{AwoCS}) gives us the simple 
solution as $A(p)=1$ and the problem reduces to solve Eq. (\ref{BwoCS}) 
with $A(p)=1$.  

In the case with the Chern-Simons term, as is seen in Eqs. (\ref{B}) and 
(\ref{A}), there does not exist such a specific gauge where the wave function 
is not renormalized.  
As far as we cannot find a self-evident reason that the Landau is still 
specific in $QED_3$ with the Chern-Simons term, it must be fair to study 
Eqs. (\ref{B}) and (\ref{A}) for various values of the gauge-fixing 
parameter $\alpha$.  

\subsection*{IV-B. Small $p$ Expansion and Perturbative Results} 

At first sight, Eqs. (\ref{B}) and (\ref{A}) seem to be dangerous in the 
limit $p \rightarrow 0$.  
But we can check that this limit is well-defined.  
In the region $p \approx 0$, Eqs.(\ref{B}) and (\ref{A}) are written as 
\begin{eqnarray}
B(p) &=& \frac{e^2}{\pi^2} \int_{0}^{\infty}dk 
            \frac{k}{A(k)^2 k^2 + B(k)^2}
            \left[ \frac{k \mu}{k^2+\mu^2} A(k) + \left( \frac{\alpha}{2 k} 
            + \frac{k}{k^2+\mu^2} \right) B(k) \right. 
\nonumber \\
     &+& \left. O(p^2) \right] \ \ , 
\label{B0} \\
A(p) &=& 1+\frac{e^2}{\pi^2} \int_{0}^{\infty}dk
           \frac{k}{A(k)^2 k^2 + B(k)^2} \left[ \frac{1}{3} \left\{ 
           \frac{\alpha}{k} - \frac{2k \mu^2}{(k^2+ \mu^2)^2} \right\} 
           A(k) \right. \nonumber \\
     &-& \left. \frac{\mu}{3k} \frac{k^2-\mu^2}{(k^2+\mu^2)^2} B(k) 
          + O(p^2) \right]  \ \ .
\label{A0}
\end{eqnarray}
(See Appendix B.)  
From Eqs. (\ref{B0}) and (\ref{A0}), we can derive the result which is 
obtained in the lowest-order of perturbation.  
By setting $A(k)=1$ and $B(k)=0$ and performing the integration on $k$, we 
have 
\begin{eqnarray}
B(0)=\frac{e^2}{2 \pi} \frac{|\mu|}{\mu} \ , \ \ 
A(0)=1 - \frac{e^2}{6 \pi} \frac{|\mu|}{\mu^2} 
+ \frac{e^2}{3 \pi^2} \frac{\alpha}{\epsilon} \ \ , 
\label{Per}
\end{eqnarray}
where $\epsilon$ is the infrared cutoff in the integration on $k$.  
It should be noticed that $B(0)$ depends on only the sign of $\mu$.  
This also may be a specific aspect in the (2+1)-dimensions.  
The dependence of $A(0)$ on $\mu$ shows that only the Landau gauge is free 
from the infrared divergence.  
On the other hand, $A(0)$ is singular at $\mu=0$ so that the theory with the 
Chern-Simons term may not be smoothly connected to the theory without the 
Chern-Simons term in the lowest-order perturbation.  

\subsection*{IV-C. Constant Approximation} 

Before proceeding to a numerical analysis, it is very useful if we can 
estimate $A(0)$ and $B(0)$ analytically even under a fairly crude 
approximation.  
The kernels of these integral equations are dumped rapidly as the integral 
variable $k$ increases so that the contribution from $k \approx 0$ is the 
most dominant one in the integrals.  
We approximate $A(k)$ and $B(k)$ by $A(0)$ and $B(0)$ in the integrals.  
We call this approximation "the $constant$ approximation".  
Of course this approximation might be too crude for our purpose and we only 
use the result as reference in the numerical analysis.  
Under this approximation, we can perform the remaining radial integration and 
have the simple algebraic equations as
\begin{eqnarray}
B(0) &=& \frac{e^2}{2\pi} \frac{1}{A(0)} (\frac{\alpha}{2}+1) \ \ \ , 
\label{Balge} \\
A(0) &=& 1+\frac{e^2}{6\pi} \frac{\alpha}{B(0)} \ \ \ ,
\label{Aalge}
\end{eqnarray}
where we have considered the case of $\mu>0$.  (See Appendix B for the details.)  
Solving the coupled algebraic equations (\ref{Balge}) and (\ref{Aalge}), we 
obtain 
\begin{eqnarray}
B(0) &=& \frac{e^2}{2\pi} + \frac{e^2}{12\pi}\alpha \ \ , 
\label{Bcnst} \\
A(0) &=& 1+\frac{2\alpha}{\alpha+6} \ \ .
\label{Acnst}
\end{eqnarray}

From Eqs.(\ref{Bcnst}) and (\ref{Acnst}), we can see that the dependence 
of $B(0)$ and $A(0)$ on the gauge-fixing parameter, the coupling 
constant and the topological mass has the following peculiar features: 
\begin{itemize}
\item[1)] Dependence on the gauge-fixing parameter \\  
$B(0)$ depends linearly on $\alpha$.  
It is suggestive that $A(0)$ is singular at $\alpha=-6$ where B(0) vanishes.  
In the Landau gauge ($\alpha=0$), $A(0)=1$ and $B(0)=e^2/2\pi$.  
$A(0)=1$ is favourable for us because $A(p)=1$ means that the Ward-Takahashi 
identity is satisfied.  
\item[2)] Dependence on the coupling constant \\
$A(0)$ does not depend on $e^2$.  
It means that the deviation of $A(0)$ from 1 is independent of the coupling 
constant.  
This is crucially different from the perturbative result given by 
Eq.(\ref{Per}) where the deviation is proportional to $e^2$.  
On the other hand, $B(0)$ is proportional to $e^2$.  
\item[3)] Dependence on the topological mass \\
We recognize that Eqs.(\ref{Bcnst}) and (\ref{Acnst}) are 
independent of the topological mass $\mu$.  
In fact, if we apply the constant approximation to the case without the 
Chern-Simons term, we obtain the same results as Eqs. (\ref{Bcnst}) and 
(\ref{Acnst}).  
It means that the amount of the explicit parity breaking in the gauge sector 
by the topological mass does not affect the dynamical mass in the fermion 
sector.   
\end{itemize}
Now we proceed to a more precise numerical evaluation in the next section.  

\section*{V. Numerical Analysis}

\subsection*{V-A. Non-trivial Solutions}

We solve the two coupled integral equations (\ref{B}) and (\ref{A}) 
numerically by using a method of iteration.  
First we substitute trial functions into $A(k)$ and $B(k)$ in the right-hand 
sides of Eqs. (\ref{B}) and (\ref{A}) and then calculate the integrals 
numerically.  
The outputs so obtained, $A(p)$ and $B(p)$, are substituted back to the 
right-hand sides until the outputs coincide with the inputs.  
Finally we obtain convergent functions $A(p)$ and $B(p)$, which satisfies 
the integral equations, if there exist any solutions in Eqs. (\ref{B}) 
and (\ref{A}).  

We obtain the non-trivial solutions for the various values of the gauge 
parameter $\alpha$.  
For $e=1.0$ and $\mu=1.0$, $A(p)$'s in the cases of $\alpha =0, 1, 
2, 3$ are shown in Fig. 1.  
The corresponding $B(p)$'s are presented in Fig. 2.  
We find a very interesting feature that $A(p)$ is fairly close to 1 in the 
Landau gauge ($\alpha=0$).  


In the case of $QED_3$ without the Chern-Simons term, $A(p)$ is exactly equal 
to 1 in the Landau gauge under the lowest ladder approximation
\footnote{Outside the lowest ladder approximation, it is known that $A(p)$ 
differs from one in the infrared.~\cite{AMMKMBR}}.~\cite{HMH}  
However, in the case of $QED_3$ with the Chern-Simons, there may be no 
apparent reason that $A(p)=1$ in the Landau gauge.  
It is surprising that the numerical calculation of so complicated integral 
equations results $A(p) \approx 1$ in the Landau gauge.  
There might be a simple reason of explaining a peculiarity of the Landau 
gauge.  


\subsection*{V-B. Dependence on the gauge parameter}

The dependence of $A(0)$ and $B(0)$ on the gauge-fixing parameter $\alpha$ 
is shown in Figs. 3 and 4.  
In the region $\alpha>0$, the numerical result is consistent with the one 
obtained in the constant approximation.


For $\alpha < 0$, the numerical iteration procedure does not converge, 
but cycles between two or more functions, none of which are solutions 
to Eqs. (\ref{B}) and (\ref{A}).  
This appears to be a manifestation of the well known "doubling route to chaos" 
frequently exhibited by non-linear iterative algorithms.  
Unfortunately this prevents us from obtaining numerical solutions for negative 
values of the gauge parameter.

A plot of the gauge invariant condensate $<\bar{\psi} \psi>$ as a function of 
$\alpha$ is helpful as a way of indicating to what extent gauge symmetry is 
broken by the bare vertex approximation.  
The condensate is defined by $<\bar{\psi} \psi>=-i\lim_{x \rightarrow 0} tr 
S'_F(x)$ where $S'_F$(x) is a propagator in real space-time coordinates.  
Using the Fourier transformation and the Wick rotation, we obtain 
\begin{eqnarray}
<\bar{\psi} \psi>= \frac{1}{\pi^2} \int^\infty_0 dk 
                   \frac{k^2 B(k)}{A(k)^2 k^2 + B(k)^2} \ \ .
\label{condensate} 
\end{eqnarray}
On the other hand, a position of the pole of the Minkowski propagator also is 
gauge invariant.  
To know the position, we have to perform an analytic continuation from our 
Euclidean results to the Minkowski ones.  
But it is a hard task because what we know in the Euclidean analysis is 
$numerical$.  
In fact, it is difficult to derive a full analytic properties from the 
numerical data.  
Instead, $B(0)/A(0)$ is usually considered as an approximated value of $p^2$ 
at the pole.  
We have shown the $\alpha$-dependence of the dynamical fermion mass 
$m_{phys}=B(0)/A(0)$ and also $<\bar{\psi} \psi>$ in Fig. 5.  
It shows that the $\alpha$-dependence may be considered to be fairly weak, 
compared with the result in the constant approximation.  


The results obtained in Secs.V-A and V-B suggest that the Landau gauge is 
still the best gauge.  
Hereafter we present mainly the results obtained in the Landau gauge.  

\subsection*{V-C. Dependence on the topological mass and the coupling constant}

Now we investigate how the dynamical fermion mass depends on the topological 
mass and the coupling constant.  
What we are most interested in is the dependence of the dynamical fermion 
mass on the topological mass of the gauge field.  
In the constant approximation, it has been shown that both $A(0)$ and $B(0)$ 
do not depend on the topological mass.  
Is this true in the more precise numerical evaluation?  

In Fig. 6 we show the dependence of $a(0)$ on the dimensionless parameter 
$\hat{\mu}$ which is defined in Sec. III.  
We can see that the deviation of $a(0)$ from 1 is less than 1 \%.  
We may say that $a(0)$ is almost equal to 1 in all region of $\hat{\mu}$.  
It means that the $\hat{\mu}$- and $e^2$-dependence of $A(0)$ is extremely 
weak.  


On the other hand, the $\hat{\mu}$-dependence of $b(0)$ is presented in Fig. 7.  
It show that though $B(0)$ is almost constant in the region 
of $\mu > e^2$, it decreases as $\mu$ does if $\mu$ is comparable to or 
smaller than $e^2$.  
The $e^2$-dependence of $B(0)$ is less explicit in Fig. 7 so that we show 
the explicit $e^2$-dependence of $B(0)$ in Fig. 8.  
$B(0)$ increases as the coupling constant becomes larger.  
The $e^2$-dependence of $B(0)$ is almost linear when $e^2$ is smaller than 
$\mu$.  
This is consistent with the constant approximation.  
But the slope is smaller than the one obtained in the constant 
approximation and the line curves downward slightly as the coupling constant 
becomes larger, which may be a non-perturbative effect.  


It should be noticed that for very small or large values of the topological 
mass compared with the coupling constant, there appears a technical difficulty 
in the numerical calculation as follows:  
In principle, we do not need to cut off the region of the energy-momentum 
integrals in Eqs. (\ref{B}) and (\ref{A}).  
But in the practical prescription of the numerical integration, the infinite 
range of the integration must be replaced by a finite one so that it is not 
avoidable to introduce cut-off parameters.   
The cut-off dependence of the results must be checked carefully.  
When the topological mass $\mu$ takes very small or large values compared 
with $e^2$, 
it has been found that the integration region must be taken wider enough to 
get reliable results.  
It needs much machine power.  
In getting our results, we have checked the absence of 
the cut-off dependence in detail in the parameter region adopted here 
($\mu/e^2=10^{-2} \sim 10^4$).  
Therefore we conclude that the behaviour of $A(0)$ and $B(0)$ mentioned 
above, especially the decreasing of $B(0)$ in the region of $\mu/e^2 < 1$, 
is not due to the cut-off.  

\section*{VI. Conclusion and Discussion}

We have investigated the dynamical mass generation of the $single$ 
two-component fermion in $QED_3$ $with$ the Chern-Simons term.  
The coupled Schwinger-Dyson equations for $A(p)$ and $B(p)$, where the full 
fermion propagator is defined by 
$S'_F(p)^{-1} = A(p) \not \hspace{-0.8mm} p-B(p)$, 
have been formulated in the lowest ladder approximation.  
When the Chern-Simons term is included, there does not exist the specific 
gauge where $A(p)$ is automatically equal to 1 as the case without the 
Chern-Simons term.  
We examine the dependence of the dynamical mass on the gauge-fixing 
parameter $\alpha$, the coupling constant $e$ and the topological mass $\mu$ 
by using the approximated analytical and also the numerical methods.  

The coupled integral equations have been solved first analytically in the 
constant approximation where $A(p)$ and $B(p)$ have been replaced by $A(0)$ 
and $B(0)$ respectively.  
We have found that $B(0)$ is proportional to $\alpha$ and also to $e^2$.  
On the other hand, $A(0)$ is independent of $e$ and singular at $\alpha=-6$ 
where $B(0)$ vanishes.  
We also have found that $A(0)$ and $B(0)$ do not depend on the topological 
mass.

Keeping these facts in mind  we have proceeded to solve the set of the 
integral equations by using the numerical method.  
The dependence of $A(0)$ and $B(0)$ on $\alpha$ and $e$ are almost consistent 
with the results derived in the constant approximation.  
$B(0)$ depends on $\alpha$ linearly.  
The $e^2$-dependence of $B(0)$ also is almost linear but the value of the 
slope becomes smaller than the one in the constant approximation as $e^2$ 
increases.  
The $e$-dependence of $A(0)$ is almost absent.  

To check to what extent gauge symmetry is spoilt by the bare vertex 
approximation, we have evaluated the gauge invariant condensate 
$<\bar{\psi} \psi>$ and also $B(0)/A(0)$.  
The result has shown that $\alpha$-dependence of them may be 
considered to be fairly weak.  

Further we have discovered some novel features in the numerical analysis.  
First we have found $A(p) \approx 1$ in the Landau gauge.  
It is not entirely obvious why the numerical evaluation of so complicated 
integral equations results that $A(0)$ is almost equal to 1 at $\alpha=0$.  
We may expect that there is any simple reason.  
It also should be noticed that the trivial $A(p)$ in the Landau gauge is 
obtained merely in the lowest ladder approximation.~\cite{AMMKMBR}  
More sophisticated approximation will be considered in subsequent works.  

Secondly we have found the strange behaviour of $A(0)$ and $B(0)$ in the 
region $\alpha <0$.  
There has appeared the doubling signal in the iterations.  
The signal is a popular phenomena in a non-linear system.  
We wonder that the signal appeared here would have any physical or 
mathematical meaning or not.  

As the third, it seems that the dynamical mass of the fermion is almost 
constant if $\mu$ is larger than $e^2$, while it decreases when $\mu$ is 
comparable to or smaller than $e^2$.  
Why it is? 
In the case of $\mu \neq 0$, we have found just one solution in our numerical 
analysis.  
On the other hand, we know that there exist two solutions in the case of 
$\mu=0$ ~\cite{HMH}; one is trivial ($B=0$) and the other one nontrivial 
($B \neq 0$).  
We have extrapolated our numerical data $B(0)/A(0)$ to $\mu=0$ by using the 
method of least squares~\cite{MLS} and obtained Fig. 9.  
The numerical evaluations in $QED_3$ without the Chern-Simons term results 
$B(0)=0.104755$ ($\alpha=0$ and $A(0)=1$).  
The value obtained by the extrapolation is $B(0)/A(0)=0.105255$ so that both 
results may be consistent with each other.  
Thus, the solution in the case of $\mu \neq 0$ seems to approach the 
non-trivial one as $\mu \rightarrow 0$.   
Of course, we should be careful before accepting this result.  
As mentioned in V-C, the numerical evaluation in the region of $\mu/e^2 <<1$ 
is very difficult technically and the behaviour of the solution in 
taking the limit $\mu \rightarrow 0$ is still not clear.  
There might appear any critical behaviour such as a bifurcation of the 
solution.  
The best way to confirm the result is to give a proof by analytical method but 
it is very difficult because the integral equations are highly non-linear.    
We are now extending the parameter region to search any more non-trivial 
structure especially in the region of $\mu/e^2<<1$.  
The result will be reported separately.  

\newpage

\section*{Appendix A: Angular Integration}

We present here the formulae for the angular integration, which are used 
to derive Eqs.(\ref{B}) and (\ref{A}) in Sec. III.  
After the Wick rotation, Eqs. (\ref{Bd3k}) and (\ref{Ad3k}) can be rewritten 
in the polar coordinates as 
\begin{eqnarray}
B(p)&=&\frac{e^2}{8\pi^2}\int_{0}^{\infty}dk
                     \frac{k^2}{A(k)^2 k^2+ B(k)^2} \left[
                     -2\mu(p^2-k^2) A(k) I_0 
    \right. \nonumber \\
    &+& \left. 2 \mu A(k) I_1 + 4 B(k) I_1 + 2\alpha B(k) I_2 \right] \ \ , 
\nonumber \\
A(p)&=&1+\frac{e^2}{8\pi^2 p^2}
                    \int_{0}^{\infty}dk\frac{k^2}{A(k)^2 k^2 + B(k)^2}
                    \left[ A(k) \left\{(p^2-k^2)^2 I_0 - I_3 \right\}
    \right. \nonumber \\
    &+& \left. 2 \mu B(k) \left\{(p^2-k^2) I_0 + I_1 \right\}
                     - \alpha A(k) \left\{(p^2-k^2)^2 I_4
                     -(p^2+k^2) I_2 \right\}\right] \ \ .
\nonumber
\end{eqnarray}
Each integrals $I_n$'s (n=0, 1, 2, 3, 4) are calculated as 
\begin{eqnarray}
I_0&=&\int_{0}^{\pi}\sin\theta d\theta\frac{1}{\{(p-k)^2+\mu^2\} (p-k)^2}
          =\frac{1}{2\mu^2pk}\ln\frac{(p+k)^2\{(p-k)^2+\mu^2\}}
                                     {(p-k)^2\{(p+k)^2+\mu^2\}} 
\ \ \ , \nonumber \\
I_1&=&\int_{0}^{\pi}\sin\theta d\theta\frac{1}{(p-k)^2+\mu^2}
          =\frac{1}{2pk}\ln \frac{(p+k)^2+\mu^2}{(p-k)^2+\mu^2} 
\ \ \ , \nonumber \\
I_2&=&\int_{0}^{\pi}\sin\theta d\theta\frac{1}{(p-k)^2}
          =\frac{1}{2pk}\ln\frac{(p+k)^2}{(p-k)^2} 
\ \ \ , \nonumber \\
I_3&=&\int_{0}^{\pi}\sin\theta d\theta\frac{(p-k)^2}{(p-k)^2+\mu^2} 
          =\frac{1}{2pk} \left\{(p+k)^2-(p-k)^2 
          - \mu^2 \ln\frac{(p+k)^2+\mu^2}{(p-k)^2+\mu^2} \right\} 
\ \ \ , \nonumber \\
I_4&=&\int_{0}^{\pi}\sin\theta d\theta\frac{1}{(p-k)^4}
          =-\frac{1}{2pk}\left\{\frac{1}{(p+k)^2}-\frac{1}{(p-k)^2} \right\}
\ \ \ . \nonumber
\end{eqnarray}

\section*{Appendix B: $p \rightarrow 0$ limit and Constant Approximation}

The formulae which are used in Sec. IV are summarized here.  
The ($p \rightarrow 0$) limit is taken by the expansion formulae as 
\begin{eqnarray}
\ln \frac{(p+k)^2}{(p-k)^2} &=& 4 \frac{p}{k} + \frac{4}{3} (\frac{p}{k})^3 
                               + O(p^5) 
\ \ \ , \nonumber \\
\ln \frac{(p+k)^2+\mu^2}{(p-k)^2+\mu^2} &=& \frac{4kp}{k^2+\mu^2} 
        + \frac{4}{3} \frac{(k^2-3\mu^2)k}{(k^2+\mu^2)^3} p^3 
        + O(p^5) 
\ \ \ . \nonumber
\end{eqnarray}

In the constant approximation of Eqs. (\ref{B}) and (\ref{A}), we replace 
the unknown functions $A(k)$ and $B(k)$ to $A(0)$ and $B(0)$.  
Setting $p=0$, we obtain 
\begin{eqnarray}
B(0) &=& \frac{e^2}{\pi^2} 
         \left[ \left\{ \mu A(0)+(\frac{\alpha}{2}+1) B(0) \right\} J_0 
      - \mu^2 (\mu A(0)+B(0)) J_1 \right] \ \ \ , 
\nonumber \\
A(0) &=& 1 + \frac{e^2}{\pi^2} 
         \left[ \frac{\alpha}{3} A(0) J_0 
              - \frac{\mu}{3} (2\mu A(0)+B(0)) J_1 
              + \frac{\mu^3}{3} (2\mu A(0)+2B(0)) J_2 \right] \ \ , 
\nonumber
\end{eqnarray}
where $J_i$'s ($i=0, 1, 2$) are given by
\begin{eqnarray}
J_0&=& \int^\infty_0 \frac{dk}{A(0)^2 k^2+B(0)^2}
    = \frac{\pi}{2A(0)B(0)} 
\ \ \ , \nonumber \\
J_1&=& \int^\infty_0 \frac{dk}{A(0)^2 k^2+B(0)^2} \frac{1}{k^2+\mu^2} 
= \frac{\pi}{2B(0)} \frac{1}{|\mu|} \frac{1}{|\mu| A(0)+B(0)} 
\ \ \ , \nonumber \\
J_2&=& \int^\infty_0 \frac{dk}{A(0)^2 k^2+B(0)^2} \frac{1}{(k^2+\mu^2)^2} 
= \frac{\pi}{4B(0)} \frac{1}{|\mu|^3} 
\frac{2|\mu|A(0)+B(0)}{(|\mu| A(0)+B(0))^2} \ \ \ .
\nonumber
\end{eqnarray}

\section*{Acknowledgment}
One of the authors (T. M.) would like to thank Prof. M. Kenmoku for his 
hospitality at the Department of Physics, Nara Women's University.

\begin{figure}
\epsfysize=7cm
\centerline{\epsfbox{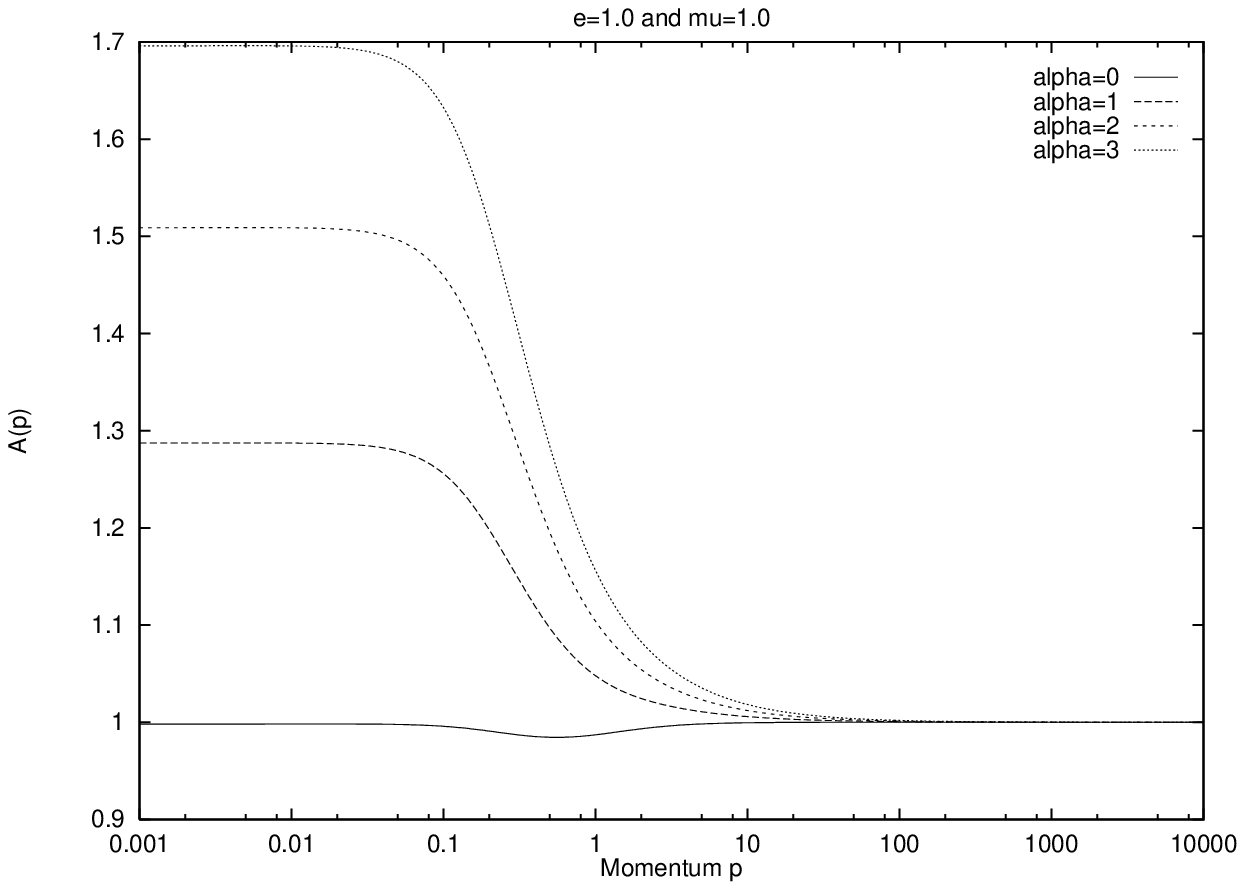}}
\caption
{The nontrivial solutions of $A(p)$ at $e=1.0$ and $\mu=1.0$
}
\end{figure}

\begin{figure}
\epsfysize=7cm
\centerline{\epsfbox{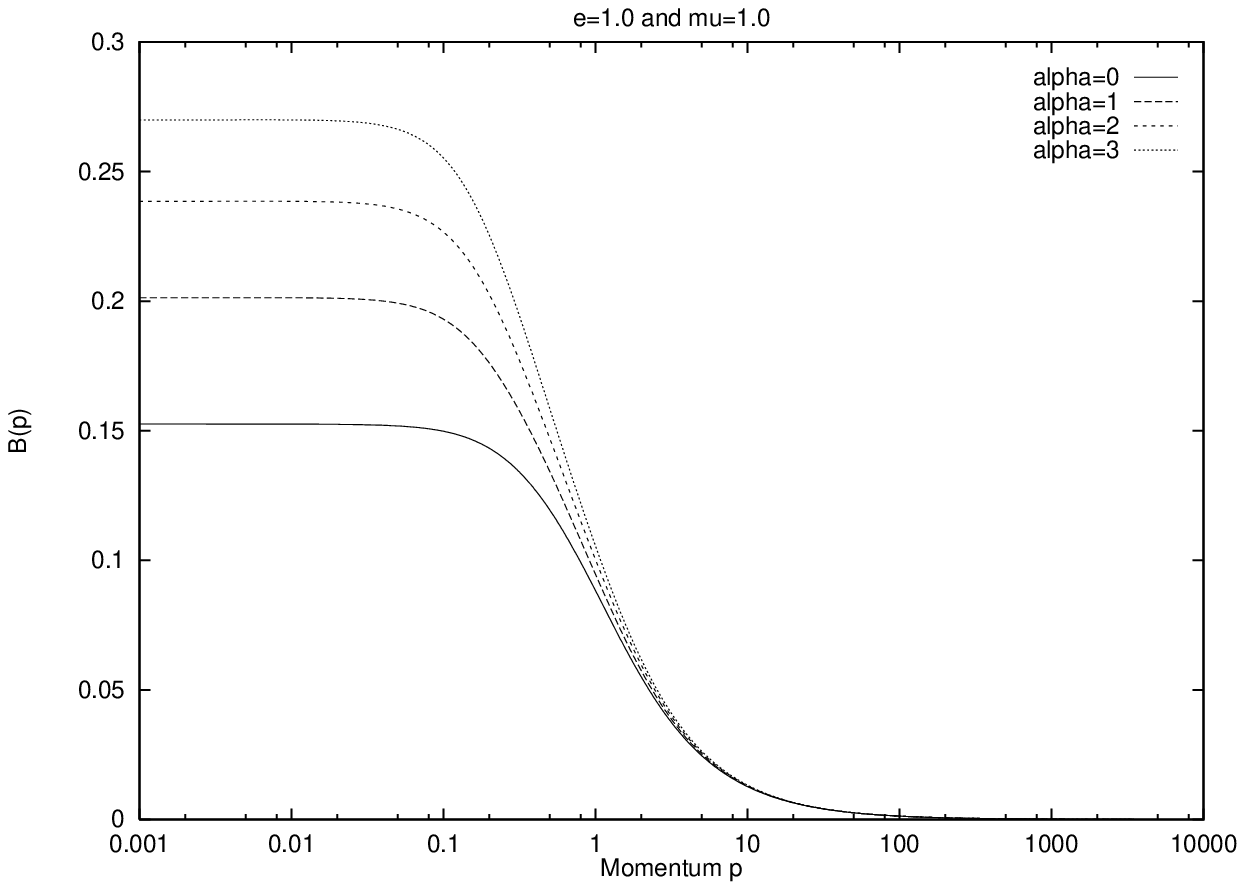}}
\caption
{The nontrivial solutions of $B(p)$ at $e=1.0$ and $\mu=1.0$
}
\end{figure}

\begin{figure}
\epsfysize=7cm
\centerline{\epsfbox{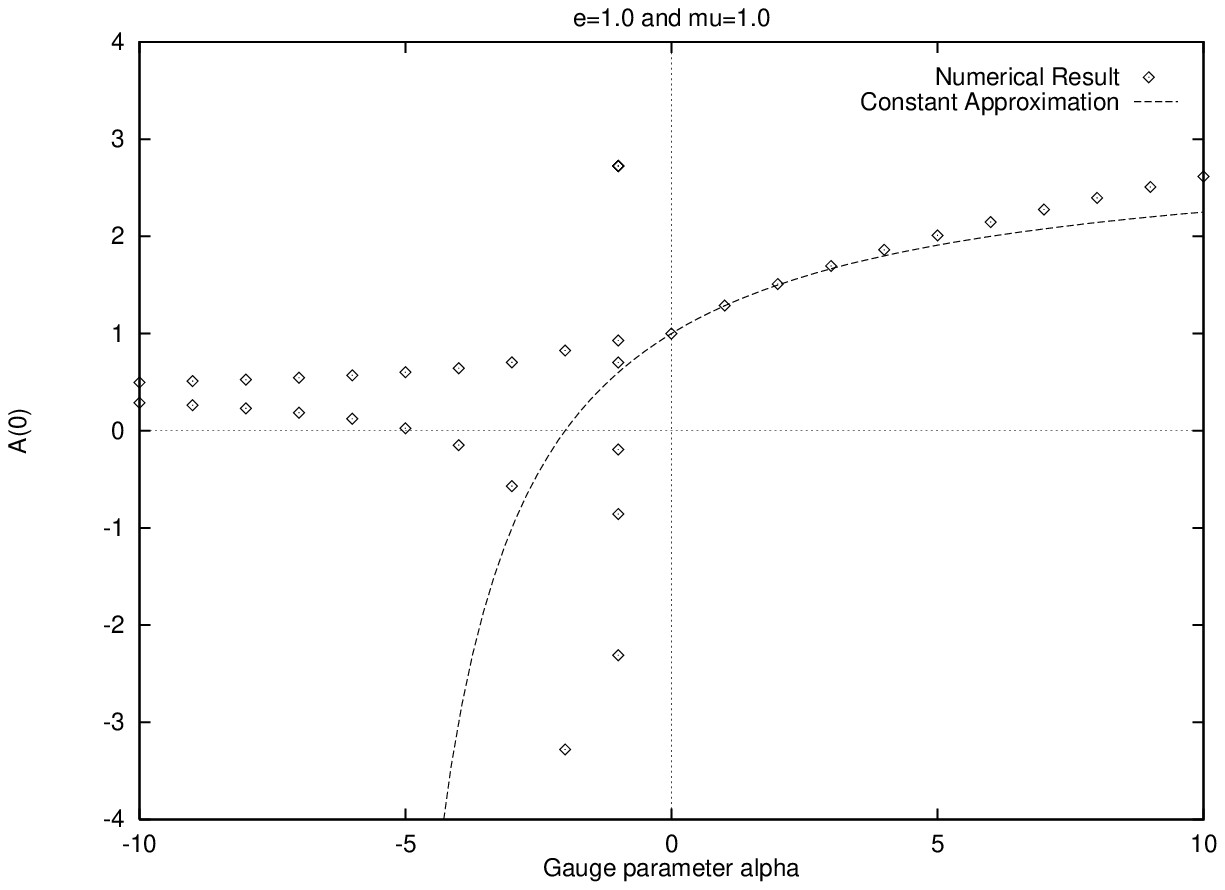}}
\caption
{The $\alpha$-dependence of $A(0)$ at $e=1.0$ and $\mu=1.0$
}
\end{figure}

\begin{figure}
\epsfysize=7cm
\centerline{\epsfbox{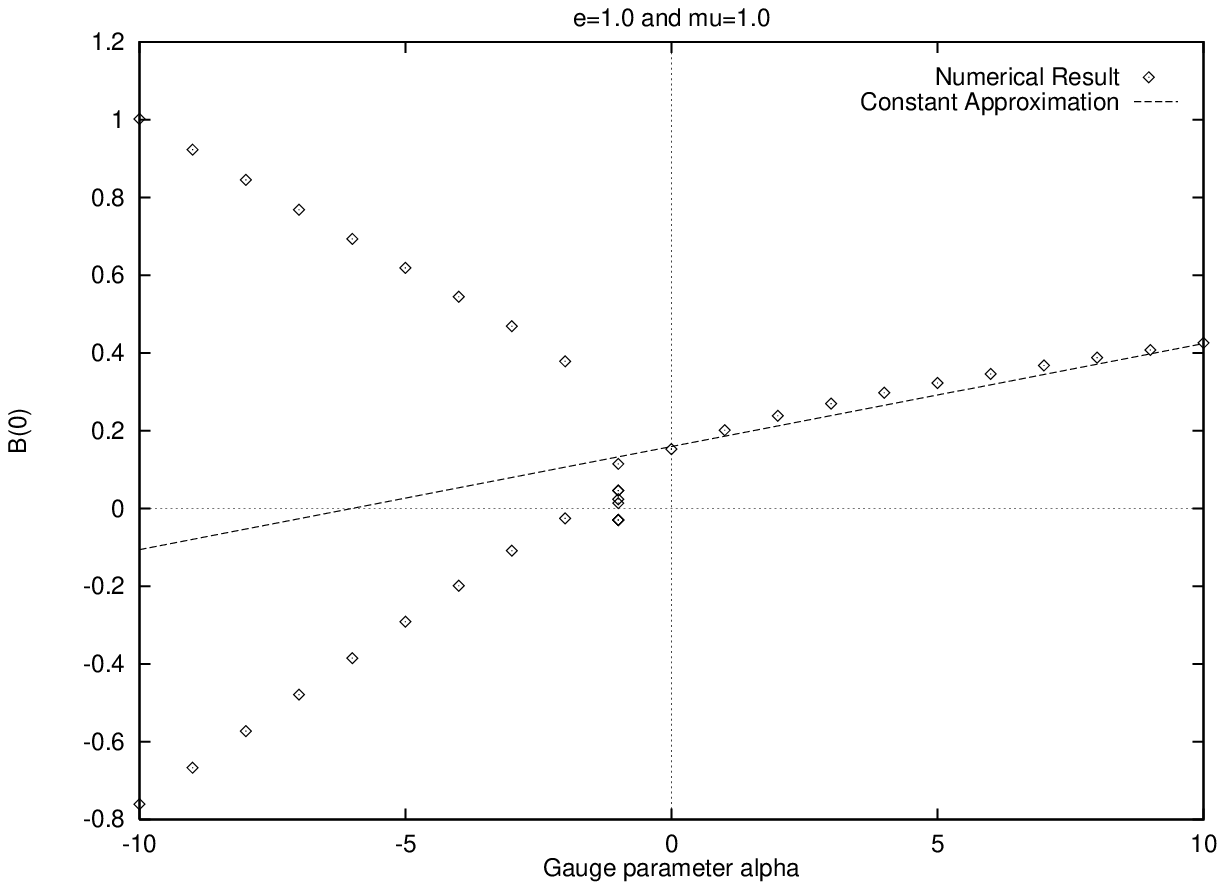}}
\caption
{The $\alpha$-dependence of $B(0)$ at $e=1.0$ and $\mu=1.0$
}
\end{figure}

\begin{figure}
\epsfysize=7cm
\centerline{\epsfbox{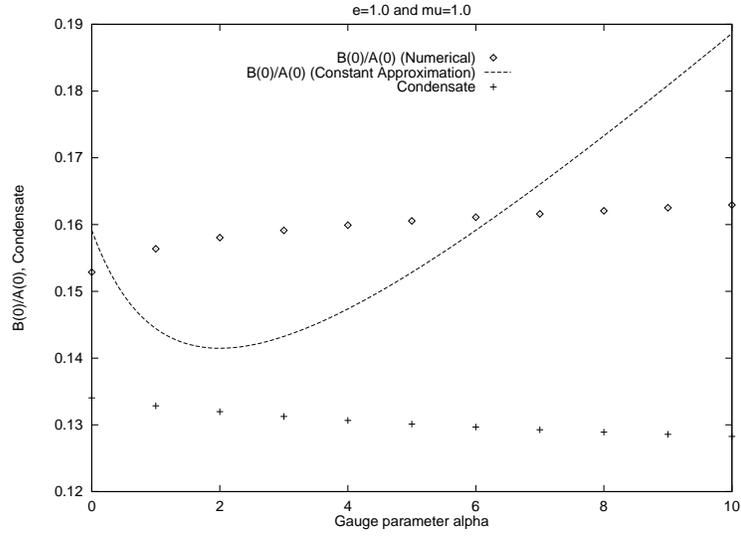}}
\caption
{The $\alpha$-dependence of $B(0)/A(0)$ and $<\bar{\psi} \psi>$ 
at $e=1.0$ and $\mu=1.0$ ($0 \leq \alpha \leq 10$)
}
\end{figure}

\begin{figure}
\epsfysize=7cm
\centerline{\epsfbox{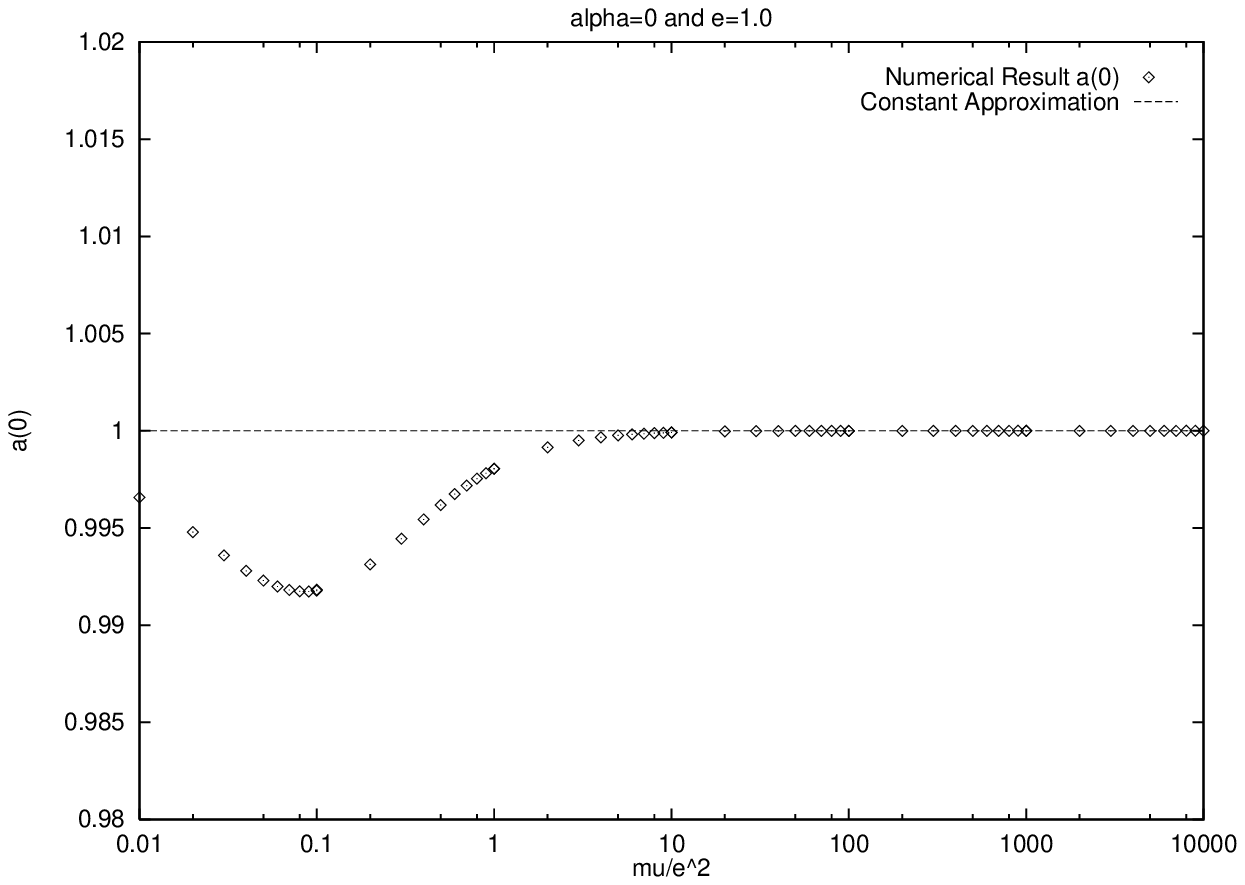}}
\caption
{The $\hat{\mu}$-dependence of $a(0)$ at $\alpha=0$ and $e=1.0$ 
}
\end{figure}

\begin{figure}
\epsfysize=7cm
\centerline{\epsfbox{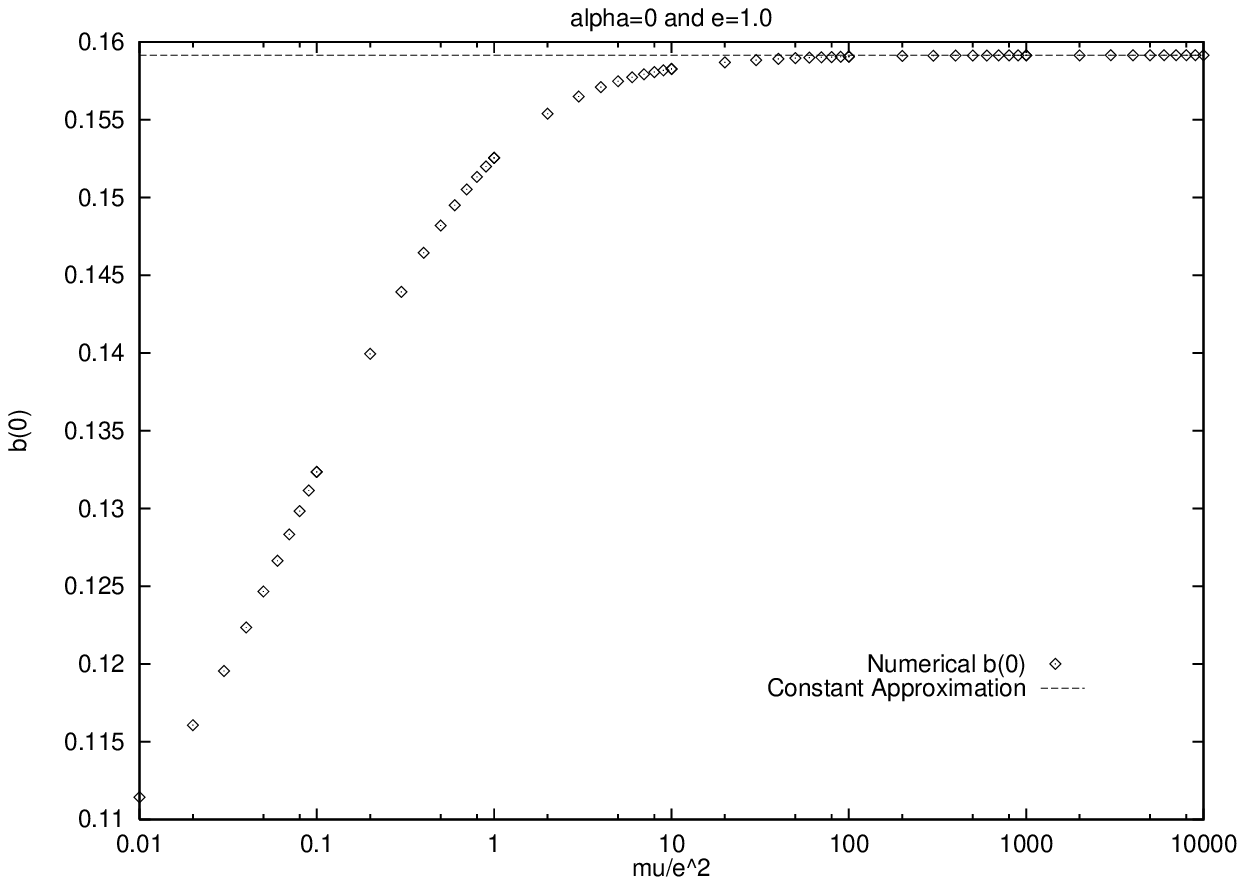}}
\caption
{The $\hat{\mu}$-dependence of $b(0)$ at $\alpha=0$ and $e=1.0$
}
\end{figure}

\begin{figure}
\epsfysize=7cm
\centerline{\epsfbox{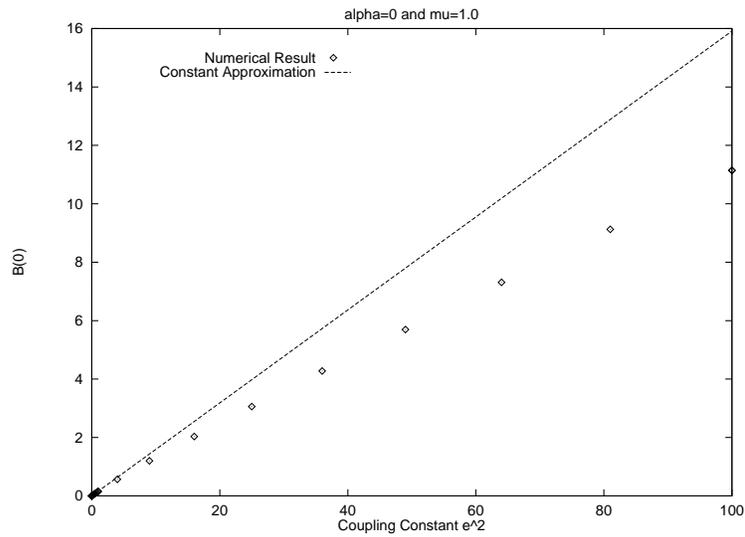}}
\caption
{The $e^2$-dependence of $B(0)$ at $\alpha=0$ and $\mu=1.0$ 
($1.0 \times 10^{-4}<e^2<100$)
}
\end{figure}

\begin{figure}
\epsfysize=7cm
\centerline{\epsfbox{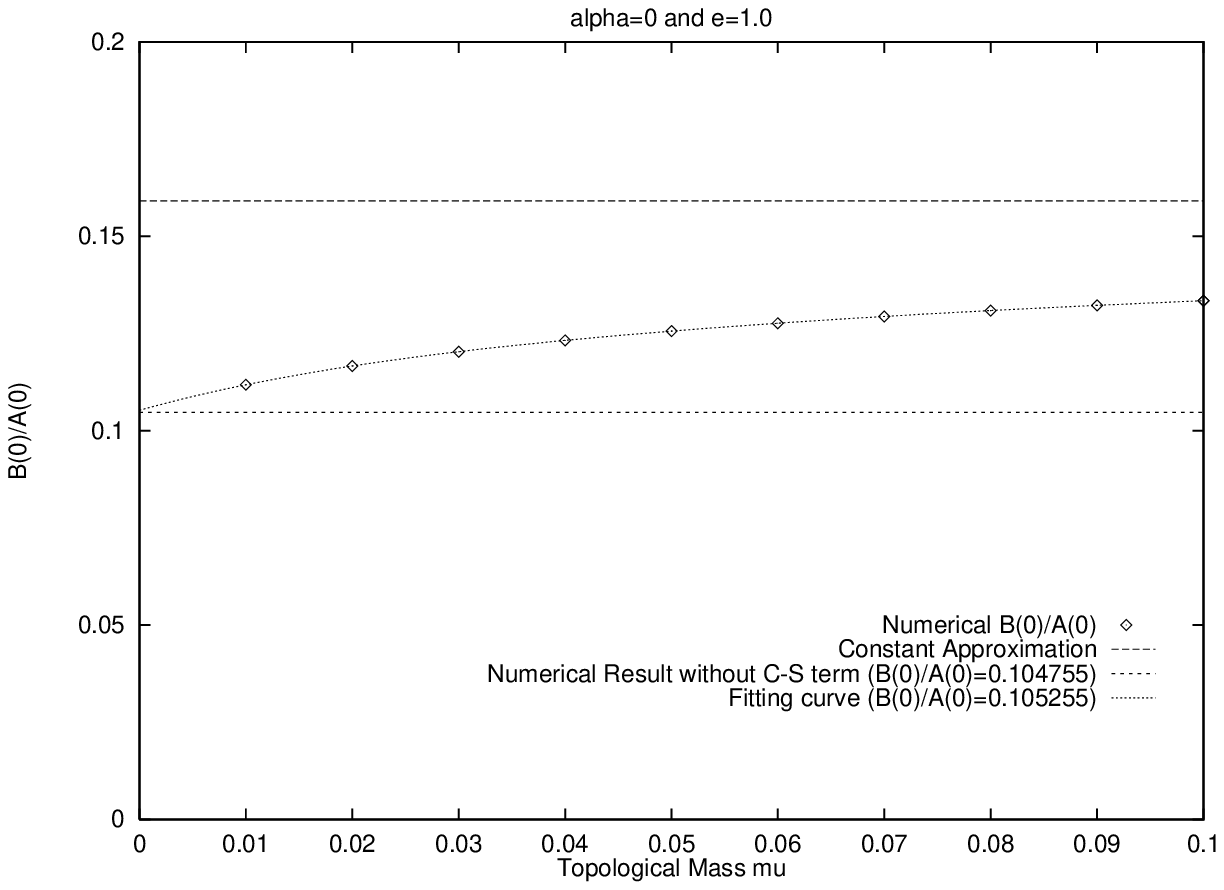}}
\caption
{The extrapolation of $B(0)/A(0)$ in the region of $\mu = 0.01 \sim 0.1$ to 
$\mu=0$ ($\alpha=0$ and $e=1.0$).  
}
\end{figure}


\begin{thebibliography}{100}
\bibitem{MLS}
         We have fitted the values of $B(0)/A(0)$ in $\mu=0.01 \sim 0.1$ by 
         using the method of least squares.  
         The fitting curve in Fig. 15 is given by the 5th-order polynomial as
         $B(0)/A(0)=0.10525482+0.76602688\mu-12.316138\mu^2
                   +144.377406\mu^3-964.29214\mu^4+2680.59064\mu^5$ .
\end{thebibliography}
\end{document}